\documentclass[preprint,12pt]{elsarticle}

\usepackage{amssymb}
\usepackage{graphicx,amsopn,subfigure}
\usepackage{algorithmic,algorithm}
\usepackage{booktabs}
\usepackage{caption2}
\DeclareMathOperator*{\argmax}{arg\,max}

\journal{Physica A}

\begin{document}

\begin{frontmatter}



\title{Advanced modularity-specialized label propagation algorithm for detecting communities in networks}


\author{X. Liu\corref{l}}
\ead{tsinllew@ai.cs.titech.ac.jp}

\author{T. Murata\corref{m}}

\cortext[l]{Corresponding author}

\address{Tokyo Institute of Technology, 2-12-1 Ookayama, Meguro, Tokyo 152-8552, Japan}

\begin{abstract}
A modularity-specialized label propagation algorithm (LPAm) for detecting network communities was recently proposed. This promising algorithm offers some desirable qualities. However, LPAm favors community divisions where all communities are similar in total degree and thus it is prone to get stuck in poor local maxima in the modularity space. To escape local maxima, we employ a multistep greedy agglomerative algorithm (MSG) that can merge multiple pairs of communities at a time. Combining LPAm and MSG, we propose an advanced modularity-specialized label propagation algorithm (LPAm+). Experiments show that LPAm+ successfully detects communities with higher modularity values than ever reported in two commonly used real-world networks. Moreover, LPAm+ offers a fair compromise between accuracy and speed.
\end{abstract}

\begin{keyword}
community detection \sep modularity \sep network \sep graph \sep clustering

\PACS 89.75.Hc \sep 05.10.-a

\end{keyword}

\end{frontmatter}

% \linenumbers


\section{Introduction}\label{int}

Detecting communities in networks has attracted a great deal of interest recently. Informally, a community is a densely connected subnetwork that is only sparsely linked to the remaining network. It is said that constructing algorithms for detecting communities is of great importance as it provides insight into the structures of real-world systems \cite{mas}.

Modularity \cite{mej2} is a scalar value that measures the quality of a particular division of a network into communities. Among various kinds of methods for detecting network communities, one that is widely used is modularity optimization \cite{san}. The modularity optimization method detects communities by searching over possible divisions of a network for one that have particularly high modularity. Since finding the ``best" community division with the highest modularity value is proven to be NP-hard \cite{ubr}, exhaustive search over all possible divisions is in general intractable. Therefore, all of the modularity optimization algorithms are based on approximate optimization.

Recently Raghavan, Albert et al. propose a label propagation algorithm (LPA) \cite{unr} for detecting network communities. This innovative and promising algorithm uses only the network structure as a guide, and can detect communities at very high speed. Barber and Clark extend LPA by relating it to modularity, and introduce a modularity-specialized LPA (LPAm) \cite{mic}. However, it is found that LPAm is prone to get stuck in poor local maxima in the modularity space \cite{mic}. To detect communities with high modularity values, we improve LPAm by driving it out of local maxima and devise a new algorithm called LPAm+ in this paper. Experiments show that LPAm+ successfully detects communities with the highest modularity values in several commonly used real-world networks.

The structure of the paper is as follows: in the next section, the definition of modularity is reviewed. In section \ref{lpa}, we give a survey of LPA and LPAm. Our algorithm is proposed in section \ref{lpamplus}. Experiments are shown in section \ref{exp}, followed by a conclusion and discussion in the last section.

\section{Modularity}
\label{mod}
To evaluate the goodness of a particular division of a network into communities, Newman introduces a measure called modularity \cite{mej2}. Consider a (undirected and unweighted) network with {\it n} nodes and {\it m} edges represented by an adjacency matrix A, whose element A{\scriptsize {\it u}{\it v}} is equal to 1 if there is an edge between nodes {\it u} and {\it v}, and 0 otherwise. The degree of a node {\it u} is denoted by {\it k{\scriptsize u}}. Suppose a particular division of the network into {\it N{\scriptsize c}} communities, such that each node {\it u} is assigned to a community {\it l{\scriptsize u}}. Modularity essentially measures the actual fraction of intra-community edges minus its expected value in a null model, where the community division is the same but connections are made randomly between nodes. Formally, modularity is defined as:
\begin{eqnarray}
Q & = & \frac{1}{2m} \sum_{u,v=1}^n ({\rm A}_{uv}-{\rm P}_{uv})\delta(l_{u},l_{v}), \label{eq1}
\end{eqnarray}
where $ {\rm P}_{uv}={k_u k_v}/2m $ is probability in the null model that an edge exists between nodes {\it u} and {\it v}, and $ \delta(i, j) $ is the Kronecker's delta. Further, a modularity matrix B is defined with elements $ {\rm B}_{uv}={\rm A}_{uv}-{\rm P}_{uv}$. Hence, modularity is expressed as:
\begin{eqnarray}
Q = \frac{1}{2m} \sum_{u,v=1}^n {\rm B}_{uv}\delta(l_{u},l_{v}).\label{eq2}
\end{eqnarray}
We can also reformulate modularity as the addition of contributions over all communities:
\begin{eqnarray}
Q = \sum_{t=1}^{N_c} \left(\frac{I_t}{m}-\left(\frac{D_t}{2m}\right)^2\right),\label{eq3}
\end{eqnarray}
where $ I_t $ is the number of intra-community edges that have both ends in community {\it t}, $D_t$ the sum over all degrees of nodes in community {\it t}.

\section{LPA}\label{lpa}
In this section, we give a survey of LPA and LPAm, which are the bases of the following discussion.

\subsection{LPA}\label{slpa}

The idea of LPA is simple \cite{unr}: initially each node in the network is assigned with a unique label, indicating the community it belongs to. At every label propagation step, each node sequentially updates its label to a new one which is the most frequent label among its neighbors. Formally, the label updating rule for node {\it x} is:
\begin{eqnarray}
l^{new}_x = {\argmax_{l}} \left(\sum_{u=1}^{n} {\rm A}_{ux}\delta(l_{u},l)\right),\label{eq4}
\end{eqnarray}
where $l_x^{new}$ indicates new label for node {\it x}. If more than one label are the most frequent ones, the new label is chosen randomly from them. The label propagation step is performed iteratively until each node has a label that is (one of) the most frequent label(s) of its neighbors. Finally communities are identified as groups of nodes bearing the same labels.

The most striking feature of LPA is its less expensive computation than what is possible so far (near linear time complexity) \cite{unr}. The weakness is that LPA is not stable: the algorithm is sensitive to the order in which node labels are updated in each step. Thus the solutions (and their corresponding modularity values) can be quite different in different runs \cite{ian}. Sometimes LPA may even end up with a trivial solution---all nodes are identified in the same community \cite{mic}.

\subsection{LPAm}\label{slpam}

Barber and Clark extend LPA by modifying the label updating rule so that modularity can be maximized, and propose a new algorithm called LPAm \cite{mic}. We can rewrite (\ref{eq2}) by separating elements regarding the label of node {\it x} from others, yielding:
\begin{eqnarray}
Q = \frac{1}{2m} \left(\sum_{u \ne x}\sum_{v \ne x} {\rm B}_{uv} \delta(l_{u},l_{v}) - {\rm B}_{xx} \right)+ \frac{1}{m} \left( \sum_{u=1}^{n} {\rm B}_{ux} \delta(l_{u},l_{x}) \right). \label{eq5}
\end{eqnarray}
When updating the label for {\it x}, by selecting a new label that maximizes the second term on the right hand side of (\ref{eq5}), we actually maximize {\it Q}. Hence, to consider updating the label for node {\it x}, the updating rule of LPAm is:
\begin{eqnarray}
l^{new}_x = {\argmax_{l}} \left(\sum_{u=1}^{n} {\rm B}_{ux}\delta(l_{u},l)\right). \label{eq6}
\end{eqnarray}
Implementing LPAm would bring about a monotone increase in modularity, hindering the trivial solution being formed. Besides, LPAm preserves the merit of high speed of LPA. But LPAm is prone to get stuck in poor local maxima in modularity space, with a similar total degree of nodes in different communities \cite{mic}. Moreover, LPAm still suffers from the weakness of instability.

\section{LPAm+}\label{lpamplus}

In this section, we first give an example in which LPAm gets stuck in a local maximum, then we introduce how to escape the local maximum and propose our improved algorithm LPAm+.

\begin{figure}
\begin{center}
\subfigure[]{\label{fig1a}
\begin{minipage}[b]{0.4\textwidth}
\includegraphics[width=1\textwidth]{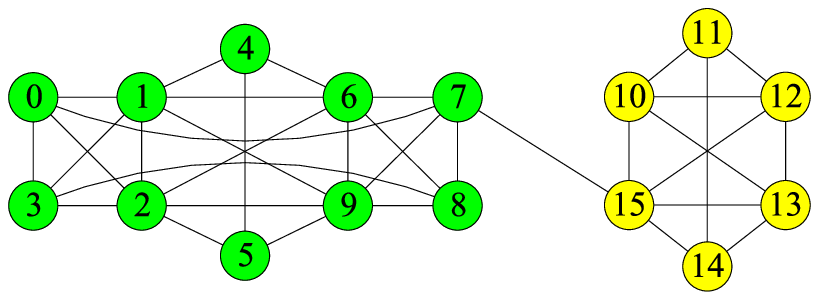}
%\caption{(a)}
\end{minipage}
}
\mbox{\hspace{1cm}}
\subfigure[]{\label{fig1b}
\begin{minipage}[b]{0.4\textwidth}
\includegraphics[width=1\textwidth]{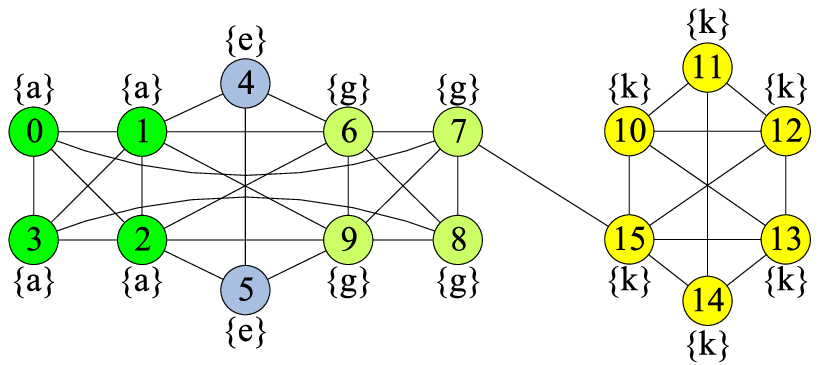}
\end{minipage}
}\\
\subfigure[]{\label{fig1c}
\begin{minipage}[b]{0.4\textwidth}
\includegraphics[width=1\textwidth]{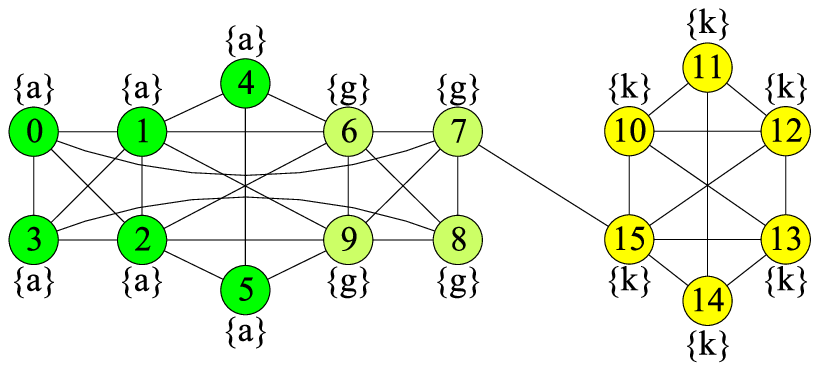}
\end{minipage}
}
\mbox{\hspace{1cm}}
\subfigure[]{\label{fig1d}
\begin{minipage}[b]{0.4\textwidth}
\includegraphics[width=1\textwidth]{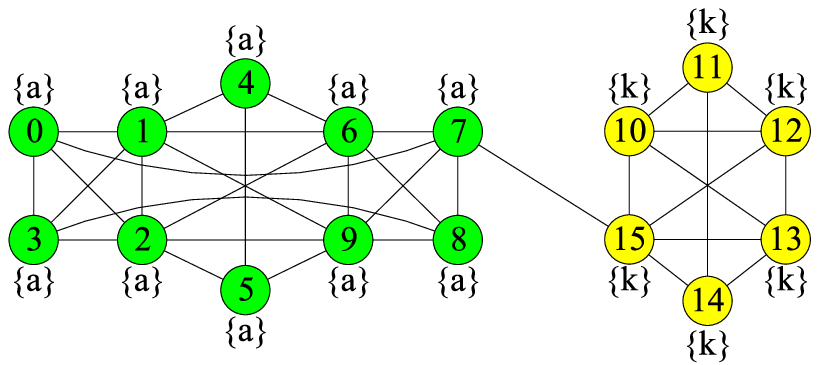}
\end{minipage}
}
\end{center}
\caption{\label{fig1}A toy network. (a) The network is intuitively divided into two communities. (b) LPAm gets stuck in the poor local maximum where the network is divided into 4 communities and the modularity is 0.399. (c) We escape the local maximum descried in (b) by merging the community labeled `a' and `e', with modularity increased by 0.008. (d) After carrying out LPAm again, we climb onto another local maximum, which is also the global maximum, with modularity increasing from 0.407 to 0.413.}
\end{figure}

Take the toy network shown in Fig. \ref{fig1a} as an example. This network is intuitively divided into two communities (painted in yellow and green colors respectively), with its modularity equal to 0.413. Feeding LPAm with this network, we obtain a division into four communities (Fig. \ref{fig1b}) and its modularity is 0.399. Evidently this division corresponds to a local maximum in the modularity space. Under the label updating rule (\ref{eq6}), LPAm favors community divisions where all communities are similar in total degree, which immediately leads to the separation of communities \{0-3\}, \{4,5\} and \{6-9\}.

To escape the local maximum, we have to get rid of the current constraint. Note that (\ref{eq6}) is a modularity maximization rule based on local structure of the network. Viewing broadly, we can adopt the greedy rule for merging communities that maximizes modularity: when LPAm gets stuck in a local maximum (no modularity gain can be achieved from further label propagation), we calculate the modularity changes for merging pairs of communities, and merge those pairs that improve modularity most. In real operation, we employ the technique used in the multistep greedy agglomerative algorithm (MSG) \cite{phi} that promotes simultaneously merging of multiple pairs of communities at a time, under the following criteria: suppose $t_1$ and $t_2$ is a pair of communities to be merged; neither $t_1$ nor $t_2$ is present in another pair inducing a higher modularity changes (see Appendix \label{app} for additional details).

After merging communities, we escape the local maximum. Then we should carry out another round of LPAm. This is analogous to climbing onto another local maximum. However, it is not guaranteed that the new local maximum we arrived at is good enough (although it is better than the previous local maximum). Hence we should repeat the above process (escaping the local maximum and climbing onto another local maximum) for many times, until no improvement of modularity can be reached. Fig. \ref{fig1b}-\ref{fig1d} give an illustration for LPAm+ working in the toy network. The pseudo-code of LPAm+ is presented in Algorithm \ref{algol}. It is clear that LPAm+ brings a monotone increase in modularity. Since the time complexity estimation of LPAm+ is a little complex, we left it to the next section.

\begin{algorithm}
\fontsize{10pt}{10pt}\selectfont
\caption{LPAm+.} \label{algol}
\begin{algorithmic}[1]
\STATE Each node is assigned with a unique label
\STATE maximize modularity by LPAm
\WHILE {$\exists$ community pair ($t_1$, $t_2$) with {\scriptsize $\Delta Q_{t_1 t_2}>0$}}
\FOR {every community pair ($t_1$,$t_2$): {\scriptsize (\hspace{0.2mm}$\Delta Q_{t_1 t_2}>0\hspace{0.2mm}) \hspace{1mm}\wedge\hspace{1mm}$[\hspace{0.5mm}$!\hspace{0.2mm}\exists\hspace{0.2mm}t:(\Delta Q_{t t_1}>\Delta Q_{t_1 t_2})\vee(\Delta Q_{t t_2}>\Delta Q_{t_1 t_2}$)\hspace{0.2mm}]}}
\STATE merge communities $t_1$ and $t_2$;
\ENDFOR
\STATE maximize modularity by LPAm
\ENDWHILE
\end{algorithmic}
%\noindent\rule[0.1\baselineskip]{\textwidth}{0.1pt}\\%[-0.1mm]
%$\Delta Q_{t_1 t_2}$ denotes the modularity change for merging community $t_1$ and $t_2$.
%{\tnote{$\Delta Q_{t_1 t_2}$ denotes the modularity change for merging community $t_1$ and $t_2$.}}
\scriptsize
 $*\hspace{2mm}\Delta Q_{t_1 t_2}$ {\footnotesize denotes the modularity change for merging community $t_1$ and $t_2$}.
\end{algorithm}

\section{Experiments}\label{exp}

We test LPAm+ in several real-world networks that are commonly used by other researchers for evaluating modularity optimization algorithms. These networks include: the karate club network (Karate Club) \cite{wwz}, the dolphin association network (Dolphins) \cite{dlu}, the network of co-purchased political books (Political Books) \cite{vkr}, the network of games between college football teams (College Football) \cite{mgi}, the network of collaborations between jazz musicians (Jazz) \cite{pgl}, the network of metabolic reactions in {\it Caenorhabdities elegans} ({\it C. elegans}) \cite{hje}, the network of email contacts at a university (E-mail) \cite{rgu}, the Pretty Good Privacy web of trust social network (PGP) \cite{mbo}, and the network of co-authorships for e-print papers posted to the condensed matter archive (Condmat2003) \cite{mej4}. As most of the researchers did, we uniformly treat all networks as undirected and unweighted, and exclude all self-loop edges. Table \ref{table1} lists the numbers of nodes and edges after preprocessing the data.

\begin{table}
\caption{\label{table1}The numbers of nodes and edges of the networks in our experiment.}
\begin{center}
%\begin{indented}
%\lineup
\fontsize{8pt}{10pt}\selectfont
\begin{tabular}{lrr}
\toprule
Network&\# of nodes&\# of edges\\
\midrule
Karate Club      &34&78\\
Dolphins         &62&159\\
Political Books  &105&441\\
College Football &115&613\\
Jazz             &198&2,742\\
{\it C. elegans}  &453&2,025\\
E-mail           &1,133&5,451\\
PGP              &10,680&24,316\\
Condmat2003      &27,519&116,181\\
\bottomrule
\end{tabular}
%\end{indented}
\end{center}
\end{table} 

We apply LPAm and LPAm+ one hundred times to each of the networks. Table \ref{table2} shows the maximal modularity, the average modularity, the standard deviation of modularity, and the average execution time collected from samples. We can see that both the maximal modularity and the average modularity obtained by LPAm+ are markedly higher than those by LPAm, consistently in all of the networks. This implies the success of our trick for escaping the local maxima. For the index of the standard deviation of modularity, we can find that LPAm+ value is significantly smaller than that of LPAm. As a matter of fact, normally the difference of modularity values between solutions of LPAm+ in different runs is within 1\%. Even in extreme cases, the difference between the worst and the best modularity values is no more than 5\%. Therefore, LPAm+ is much more stable than LPAm.

\begin{table}
\caption{\label{table2}Comparisons between LPAm and LPAm+. Values are collected from one hundred runs for each network. $Q_{max}$ denotes the maximal modularity value, $Q_{avg}$ the average modularity value, $\sigma$ the standard deviation of the modularity value, and {\it t} the average execution time (in seconds, on a PC with Intel Core 2 Duo CPU @ 2.53GHz).}
\begin{center}
\fontsize{8pt}{10pt}\selectfont
\begin{tabular}{lrrrrrrrrr}
  \toprule
  &\multicolumn{4}{c}{LPAm}&&\multicolumn{4}{c}{LPAm+}\\
%  &\centre{4}{LPAm}&\centre{4}{LPAm+}\\
%  \ns
%  Network&\crule{4}&\crule{4}\\
%  &$Q_{max}$&$Q_{avg}$&\0\0\0\0$\sigma$&$t(sec.)$&$Q_{max}$&$Q_{avg}$&\0\0\0\0$\sigma$&\0$t(sec.)$\\
  \cmidrule{2-5}\cmidrule{7-10}
  Network&$Q_{max}$&$Q_{avg}$&$\sigma$&$t$&&$Q_{max}$&$Q_{avg}$&$\sigma$&$t$\\
  \midrule
Karate Club      &0.399&0.352&0.0277&0.009&&0.420&0.418&0.0061&0.014\\
Dolphins         &0.516&0.495&0.0076&0.019&&0.529&0.523&0.0023&0.034\\
Political Books  &0.522&0.493&0.0199&0.048&&0.527&0.527&0.0011&0.088\\
College Football &0.604&0.579&0.0182&0.049&&0.605&0.604&0.0018&0.080\\
Jazz             &0.445&0.436&0.0092&0.229&&0.445&0.444&0.0013&0.368\\
{\it C. elegans} &0.409&0.379&0.0138&0.354&&0.452&0.441&0.0045&1.247\\
E-mail           &0.537&0.496&0.0155&1.097&&0.582&0.576&0.0028&3.589\\
PGP              &0.726&0.705&0.0085&5.396&&0.884&0.882&0.0009&114.221\\
Condmat2003      &0.582&0.568&0.0036&31.952&&0.755&0.751&0.0012&461.599\\
  \bottomrule
\end{tabular}
\end{center}
\end{table} 

\begin{figure}
\begin{center}
\includegraphics[width=3.0in]{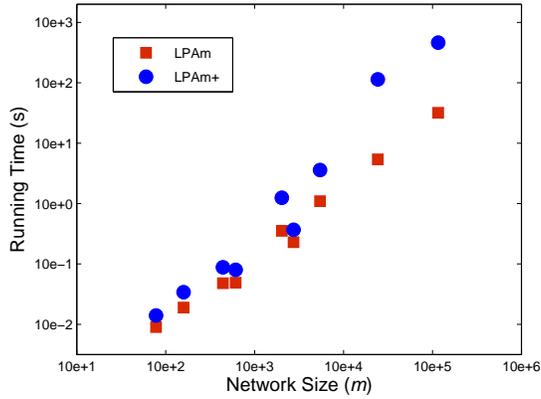}
\end{center}
\caption{\label{fig2}Comparison of running time for LPAm and LPAm+ in networks of different sizes (on a PC with Intel Core 2 Duo CPU @ 2.53GHz).}
\end{figure}

Fig. \ref{fig2} portrays the running time of LPA and LPAm+ in networks of different sizes. In the following, we give a time complexity analysis for LPAm+. On the one hand, one step of label propagation in LPAm costs O({\it m}) time \cite{mic}, so the time complexity of LPAm is O({\it rm}), where {\it r} is the number of label propagation steps required to reach a local maximum in modularity space. On the other hand, one round of merging pairs of communities that corresponds to the for-loop in Algorithm \ref{algol} requires a time of O($m{\rm log}n$) \cite{phi}. Let $h$ denote the number of needed iterations for the while-loop in Algorithm \ref{algol}. The overall time of LPAm+ can be written as O($rm$)+$h$(O($m{\rm log}n$)+O($rm$)).

An exact estimation of $h$ is not possible, as it depends on the quality of the intermediate solution obtained by LPAm. Suppose $d$ is defined as the depth of the dendrogram describing the community structure. The number of merging rounds for a single MSG algorithm (the step width is set to be $+\infty$) would be $d$. The number of merging rounds in our algorithm, namely $h$, seems a little obscure, since LPAm is performed after each merging rounds. However, note that only two cases can happen during the label propagation process in LPAm: some communities disappear and the remaining communities exchange parts of their nodes with each other.\footnote{In theory, there are three cases during the label propagation process in LPAm: 1) existent communities disappear (this situation happens when all nodes of a community select the labels of other communities as their new labels); 2) communities exchange part of their nodes with each other (this situation happens when a part of nodes of one community select the labels of other communities as their new labels); 3) new communities appear (this situation happens when some nodes select unused labels as their new labels). But case 3) never happens in practice.} Hence we can safely arrive at: $h{\sim}d$. In Table \ref{table3}, we list the true values of $h$ when LPAm+ is applied to the various networks mentioned above.

As for $r$, it is still not very well understood. In \cite{unr}, the authors suggests that the number of label propagation steps required for LPA algorithm to converge is independent to the number of nodes, and after 5 steps 95\% of the nodes can get their ``right" labels. We show the actual values of $r$ obtained from running LPAm+ in real-world networks in Table \ref{table3}. It seems that $r$ is bounded by a small constant. Therefore, $r$=o(${\log}n$).

Taken all together, in a hierarchical network where $d{\sim}{\log}n$, LPAm+ requires an overall time of O($m{\log}^{2}n$). This scaling is the same as MSG \cite{phi} and the classical greedy agglomerative algorithm \cite{aar}.

\begin{table}
\caption{\label{table3}The average number of label propagation steps required for the embedded LPAm to converge, denoted by $r$, and the number of iterations for the while-loop, denoted by $h$, when LPAm+ is applied to real-world networks. Values are averaged over one hundred runs in each of the networks. The uncertainty of the final digit, calculated as the standard error of the mean, is shown parenthetically.}
\begin{center}
%\begin{indented}
%\lineup
%\item[]
\fontsize{8pt}{10pt}\selectfont
\begin{tabular}{lrr}
\toprule
Network&$r$&$h$\\
\midrule
Karate Club      &6.13(6)&1.02(1)\\
Dolphins         &5.66(4)&1.71(4)\\
Political Books  &6.52(6)&1.98(2)\\
College Football &5.45(4)&1.02(1)\\
Jazz             &6.93(9)&1.54(4)\\
{\it C. elegans} &6.17(5)&6.95(9)\\
E-mail           &7.11(6)&6.56(8)\\
PGP              &4.61(1)&73.8(2)\\
Condmat2003      &5.59(2)&55.9(7)\\
\bottomrule
\end{tabular}
%\end{indented}
\end{center}
\end{table} 

To compare the performance of LPAm+ with other algorithms, in Table \ref{table4} we include the (maximal) modularity values obtained by LPAm+ and by many previously published methods in these networks. These methods are, in order, the hybrid algorithm of MSG algorithm in combination with node moving refinement algorithm proposed by Schuetz and Caflisch (MSG-VM) \cite{phi}, the hybrid algorithm of single-step greedy agglomerative algorithm by significance in combination with multilevel node moving refinement algorithm advanced by Noack and Rotta (SS-ML) \cite{and}, the greedy agglomerative algorithm put forward by Clauset, Newman and Moore (Greedy) \cite{aar}, the mathematical programming approach proposed by Agarwal and Kempe (VP/LP) \cite{gag}, the extremal optimization algorithm introduced by Duch and Arenas (EO) \cite{jor}, the simulated annealing implementation proposed by Guimer\`a and Amaral (SA) \cite{rog}, and the spectral optimization method suggested by Newman (SO) \cite{mej1}.

\begin{table}
\caption{\label{table4} The (maximal) modularities obtained by LPAm+ and many previously published methods.}
%\begin{flushleft}
\begin{center}
%\begin{indented}
%\lineup
%\item[]
\fontsize{8pt}{10pt}\selectfont
\begin{tabular}{lrrrrrrrrr}
\toprule
%Network&LPAm+&MSG-VM\cite{phi}&SS-ML\cite{and}&Greedy\cite{aar}&VP/LP\cite{gag}&EO\cite{jor}&SA\cite{rog}&Spectral\cite{mej1}\\
Network&LPAm+&MSG-VM&SS-ML&Greedy&VP/LP&EO&SA&SO\\
%&&\cite{phi}&\cite{and}&\cite{aar}&\cite{gag}&\cite{jor}&\cite{rog}&\cite{mej1}\\
\midrule
Karate Club	     &0.420&0.398&0.420&0.381&0.420&0.419&0.420&0.419\\
Dolphins	     &0.529&-&0.528&-&0.529&-&0.528&0.489\\
Political Books	 &0.527&-&0.527&-&0.527&-&0.527&0.399\\
College Football &0.605&0.603&0.600&0.556&0.605&-&0.605&0.602\\
Jazz             &0.445&0.445&0.445&0.439&0.445&0.445&0.445&0.442\\
{\it C. elegans} &0.452&0.450&0.446&0.412&0.450&0.434&0.450&0.435\\
E-mail           &0.582&0.575&0.577&0.503&0.579&0.574&0.579&0.572\\
PGP              &0.884&0.878&0.884&0.849&-&0.846&-&0.855\\
Condmat2003      &0.755&0.748&0.814&0.661&-&0.679&-&0.723\\
\bottomrule
\end{tabular}
%\end{indented}
\end{center}
%\end{flushleft}
\end{table} 

To make it clearer, in Table \ref{table5}, we summarize the best solutions obtained by LPAm+ and the ones with the highest modularity values ever reported in these networks. It is found that, for eight of the nine networks considered here (Karate Club, Dolphins, Political Books, College Football, Jazz, {\it C. elegans}, E-mail and PGP), LPAm+ finds the highest modularity values. Especially, for two networks ({\it C. elegans} and E-mail), LPAm+ finds modularity values higher than previously published. Only in the Condmat2003 network, LPAm+ is outperformed by SS-ML algorithm. It is interesting to note that SS-ML, which employs the single-step greedy agglomerative algorithm followed by the multilevel node moving refinement algorithm\footnote {This algorithm is designed to improve modularity by {\textquotedblleft}adjusting{\textquotedblright} misplaced nodes.}, achieves much higher modularity value than other algorithms in this network. In \cite{and}, the devisers of SS-ML argue that MSG algorithm is generally less effective than the single-step greedy agglomerative algorithm. Perhaps the reason that LPAm+ does not work well in the Condmat2003 network is that its component MSG, with a too aggressive strategy, diverts the algorithm to a suboptimal portion of the solution space. It is also worthwhile to note that that the VP/LP \cite{gag} and SA \cite{rog} algorithms can as well find the highest modularity values in some of the networks. But they are computationally much expensive and do not scale to larger networks like PGP and Condmat2003. Therefore, though not as fast as LPAm which is noted for its speed, LPAm+ offers a fair compromise between accuracy and speed.

\begin{table}
\caption{\label{table5}Comparison between the solution with maximal value of modularity obtained by LPAm+ and the one with the highest modularity ever reported. {\it N{\scriptsize c}} is the number of detected communities. {\it Q} denotes the modularity value. Sources indicate the referenced papers where we collected the data.}
\begin{center}
\fontsize{8pt}{10pt}\selectfont
\begin{tabular}{lrrrrrr}
  \toprule
  &\multicolumn{2}{c}{LPAm+}&&\multicolumn{3}{c}{Published Algorithms}\\
%&\centre{2}{LPAm+}&&\centre{3}{Published Algorithms}\\
%\ns
%Network&\crule{2}&&\crule{3}\\
%&$N_c$&$Q$&&$N_c$&$Q$&Source\\
  \cmidrule{2-3}\cmidrule{5-7}
  Network&$N_c$&$Q$&&$N_c$&$Q$&Sources\\
  \midrule
Karate Club      &4&0.420&&4&0.420&\cite{and},\cite{gag},\cite{ame},\cite{zhe}\\
Dolphins         &5&0.529&&5&0.529&\cite{gag},\cite{gxu}\\
Political Books  &5&0.527&&5&0.527&\cite{and},\cite{gag}\\
College Football &10&0.605&&10&0.605&\cite{gag},\cite{zhe}\\
Jazz             &4&0.445&&4(5)&0.445&\cite{phi},\cite{and},\cite{gag},\cite{jor}\\
{\it C. elegans} &9&0.452&&11&0.450&\cite{phi},\cite{gag}\\
E-mail           &10&0.582&&11&0.579&\cite{gag}\\
PGP              &99&0.884&&93&0.884&\cite{and}\\
Condmat2003      &72&0.755&&76&0.814&\cite{and}\\
\bottomrule
\end{tabular}
\end{center}
\end{table}

\section{Conclusion and discussion}\label{con}

In this paper, we introduce a new community detection algorithm LPAm+ based on the previously proposed algorithm LPAm. The main idea is that we try to drive LPAm out of local maxima and hereby employ MSG to merge pairs of communities which are similar in total degree. Experiments show that LPAm+ improves LPAm in terms of modularity of the detected communities, with extra computational time. Besides, LPAm+ is more stable than LPAm. Compared with other algorithms, LPAm+ distinguishes itself by its accuracy (measured by modularity) while preserving relatively high speed. The fact that LPAm+ detects the highest modularity values in almost all of the test networks is impressive.

It should be noted that the speed of LPAm+ can still be substantially improved. First, when updating the label for a node in LPAm, candidates of the new label can be safely confined to the labels of the neighbors of that node and an unused label \cite{mic} (further, we find through experiments that unused label is never selected as a new label). In light of this, we can only update the labels of nodes whose neighbors had a label change. This means only a few labels need to be updated after most of the other labels are fixed. Hence the speed of LPAm can be dramatically increased. Second, it is possible to introduce a threshold and then stop LPAm as soon as the modularity gain from the latest label propagation step does not exceed this threshold. Although these two heuristics have little influence on the final modularity value, the computational time can be reduced to a great extent (the time complexity of the algorithm remains the same, since the order of the number of iterations for the while-loop is unchanged). For example, if we apply these two heuristics (the threshold is set to be 0.00001) in the Condmat2003 network, the running time is considerably reduced from 461.599s to 96.1s, with modularity dropped by only 0.013 (based on an average value).

It is also interesting to note that MSG is not the only means to drive LPAm out of local maxima. After this work is done, we are informed that Blondel et al. use a reduction method (communities are reduced into nodes) \cite{are} to escape the local maxima involved in another algorithm different from LPAm, and propose a two-phase community detection algorithm \cite{vin}. It seems that such reduction method can also be used to drive LPAm out of local maxima.

Another important issue is that the solutions of LPAm+ in different runs, though give similar high modularity values, are not distinct in their compositions. This phenomenon is more obvious in large-scale networks. A very recent paper \cite{ben} discusses the origin of this problem. How to make the algorithm more deterministic is left for our future work.

Overall, the presented LPAm+ algorithm is a suitable choice for analyzing community structures in networks.

\section*{Acknowledgement}

The authors gratefully thank Dr. Michael Barber for valuable comments and helpful suggestions. The authors are thankful to Prof. Alex Arenas, Prof. Mark Newman and Prof. Albert-L\'aszl\'o Barab\'asi for providing network data.



\appendix

\section{The label updating rule for driving LPAm out of local maxima}
\label{app}

The label updating rule of LPAm (\ref{eq6}) can be rewritten as:
\begin{eqnarray}
l^{new}_x & = & {\argmax_{l}} \left( \sum_{u=1}^{n} \left( {\rm A}_{ux}-{\rm P}_{ux} \right) \delta(l_{u},l) \right) \label{aeq1}\\
          & = & {\argmax_{l}} \left( \sum_{u=1}^{n} {\rm A}_{ux}\delta(l_{u},l)-\frac{k_x}{2m} \sum_{u=1}^{n}k_u \delta(l_{u},l) \right) \label{aeq2}\\
          & = & {\argmax_{l}} \left( \sum_{u=1}^{n} {\rm A}_{ux}\delta(l_{u},l)-\frac{k_x D_l}{2m} \right),\label{aeq3}
\end{eqnarray}
The first term in (\ref{aeq3}) is equal to the number of {$ x $}'s neighbors labeled {$ l $}, and the second term is the product of $ k_x/2m $ ($ k_x $ denotes the degree of $ x $) and the sum of degrees of nodes labeled $ l $ ($ D_l $).

As shown in section \ref{lpamplus}, the toy network is divided into four communities by LPAm, with communities \{0-3\}, \{4,5\} and \{6-9\} being separated (Fig. \ref{fig1b}). This division corresponds to a local maximum. But what is the reason? In further analysis, we find that node 4 has one neighbor labeled `a', `e', and `g' respectively, and the sum of degrees of nodes labeled `a' or `g' is large while the sum of degrees of nodes labeled `e' is small. Consider that node 4 is being updated. For the choice of the candidate new label $l$ as `a', `e' or `g', the value of $\sum_{u=1}^{n} {\rm A}_{u4}\delta(l_{u},l)$ that is the first term of (\ref{aeq3}) would all amount to 1. Yet the value of the second term of (\ref{aeq3}) $k_4d_l/2m$ would be smaller for the choice of $l$ as `e' than as `a' or `g'. According to the updating rule (\ref{aeq3}), node 4 would keep its label unchanged and still select `e' as the new label. Similar case is applied to node 5: it would also stick to label `e' when updated. Consequently, under the current updating rule, neither node 4 nor 5 is willing to give in first. Suppose we disregard the current updating rule, and change the label of node 4 from `e' to `a' forcedly. Though this will bring about a temporary decrease in modularity, it is reasonable to expect a greater reward from the subsequent label updation for node 5 and other nodes. Conceptually, we call nodes like 4 and 5 that block the system from further progressing as stubborn nodes. It is these stubborn nodes that results in LPAm getting stuck in local maxima.

To escape the local maxima, we should attempt to let one or more of the stubborn nodes make a compromise to break down the blocked situation. Suppose $ i_1,\ldots ,i_k $ is a set of stubborn nodes labeled $ l_{(i_1,\ldots ,i_k)} $. Our trick is to keep $ i_1,\ldots ,i_k $ holding the same label and update it (let $ i_1,\ldots ,i_k $ make a compromise at the same time). Treating the labels of $ i_1,\ldots ,i_k $ separately and rewriting (\ref{eq2}), we have:
\begin{eqnarray}
Q & = & \frac{1}{2m} \left( \sum_{u \notin \{i_1,\ldots,i_k\}} \sum_{v \notin \{i_1,\ldots,i_k\} } {\rm B}_{uv} \delta(l_{u},l_{v}) - \sum_{u\in\{i_1,\ldots,i_k\}} \sum_{v\in\{i_1,\ldots,i_k\}} {\rm B}_{uv} \right) \nonumber \\
  &   & + \frac{1}{m} \left( \sum_{u=1}^{n} \sum_{v\in\{i_1,\ldots,i_k\}} {\rm B}_{uv}\delta(l_{u},l_{(i_1,\ldots ,i_k)}) \right). \label{aeq4}
\end{eqnarray}
The first term on the right hand side of (\ref{aeq4}) are independent of the label of $ i_1,\ldots ,i_k $. Hence the label updating rule to jump out of the local maxima is:
\begin{eqnarray}
l^{new}_{(i_1,\ldots ,i_k)} = \argmax_{l} \left( \sum_{u=1}^{n} \sum_{v\in\{i_1,\ldots,i_k\}} {\rm B}_{uv}\delta(l_{u},l) \right). \label{aeq5}
\end{eqnarray}
If $ l^{new}_{(i_1,\ldots ,i_k)} \not= l_{(i_1,\ldots ,i_k)} $, the change of label for this set of stubborn nodes from $l_{(i_1,\ldots ,i_k)}$ to $l^{new}_{(i_1,\ldots ,i_k)}$ is in effect equivalent to merging the community pairs labeled $l_{(i_1,\ldots ,i_k)}$ and $l^{new}_{(i_1,\ldots ,i_k)}$. In real operation, instead of identifying the stubborn nodes and then updating their label according to (\ref{aeq5}), we directly merge a pair of communities, choosing the one that result in the greatest increase in modularity.

It is often the case that there are several sets of stubborn nodes that block the system from progressing. Of course, we can merge them pair after pair. To enhance the efficiency, we adopt the technique used in MSG \cite{phi} that promotes simultaneously merging of multiple pairs of communities at a time. The implementation detail is also discussed in \cite{phi}.

%% References

%\section*{References}
\bibliographystyle{elsarticle-num}
\bibliography{references}

\begin{thebibliography}{10}
\expandafter\ifx\csname url\endcsname\relax
  \def\url#1{\texttt{#1}}\fi
\expandafter\ifx\csname urlprefix\endcsname\relax\def\urlprefix{URL }\fi
\expandafter\ifx\csname href\endcsname\relax
  \def\href#1#2{#2} \def\path#1{#1}\fi

\bibitem{mas}
M.~A. Porter, J.~P. Onnela, P.~J. Mucha, Communities in networks, Not. Amer.
  Math. Soc. 56 (2009) 1082--1097.

\bibitem{mej2}
M.~E.~J. Newman, M.~Girvan, Finding and evaluating community structure in
  networks, Phys. Rev. E 69 (2004) 026113.

\bibitem{san}
S.~Fortunato, C.~Castellano, Community structure in graphs (2007).
\newblock \href {http://arxiv.org/abs/0712.2716} {\path{arXiv:0712.2716}}.

\bibitem{ubr}
U.~Brandes, D.~Delling, M.~Gaertler, R.~Gorke, M.~Hoefer, Z.~Nikoloski,
  D.~Wagner, On modularity clustering, IEEE Transactions on Knowledge and Data
  Engineering 20 (2008) 172--188.

\bibitem{unr}
U.~N. Raghavan, R.~Albert, S.~Kumara, Near linear time algorithm to detect
  community structures in large-scale networks, Phys. Rev. E 76 (2007) 036106.

\bibitem{mic}
M.~J. Barber, J.~W. Clark, Detecting network communities by propagating labels
  under constraints, Phys. Rev. E 80 (2009) 026129.

\bibitem{ian}
I.~X.~Y. Leung, P.~Hui, P.~Li\`o, J.~Crowcroft, Towards real-time community
  detection in large networks, Phys. Rev. E 79 (2009) 066107.

\bibitem{phi}
P.~Schuetz, A.~Caflisch, Efficient modularity optimization by multistep greedy
  algorithm and vertex refinement, Phys. Rev. E 77 (2008) 046112.

\bibitem{wwz}
W.~W. Zachary, An information flow model for conflict and fission in small
  groups, Journal of Anthropological Research 33 (1977) 452--473.

\bibitem{dlu}
D.~Lusseau, K.~Schneider, O.~J. Boisseau, P.~Haase, E.~Slooten, S.~M. Dawson,
  The bottlenose dolphin community of doubtful sound features a large
  proportion of long-lasting associations, Behavioral Ecology and Sociobiology
  54 (2003) 396--405.

\bibitem{vkr}
V.~Krebs, \href{http://www.orgnet.com/}{A network of co-purchased books about
  us politics sold by the online bookseller amazon.com.} (2008).
\newline\urlprefix\url{http://www.orgnet.com/}

\bibitem{mgi}
M.~Girvan, M.~E.~J. Newman, Community structure in social and biological
  networks, Proc. Natl. Acad. Sci. USA 99 (2002) 7821--7826.

\bibitem{pgl}
P.~Gleiser, L.~Danon, Community structure in jazz, Advances in Complex Systems
  6 (2003) 565.

\bibitem{hje}
H.~Jeong, B.~Tombor, R.~Albert, Z.~N. Oltvai, A.~L. Barab\'asi, The large-scale
  organization of metabolic networks, Nature 407 (2000) 651--654.

\bibitem{rgu}
R.~Guimer\`a, L.~Danon, D.~A. Guilera, F.~Giralt, A.~Arenas, Self-similar
  community structure in a network of human interactions, Phys. Rev. E 68
  (2003) 065103.

\bibitem{mbo}
M.~Bogu{\~n}{\'a}, R.~Pastor-Satorras, A.~D\'iaz-Guilera, A.~Arenas, Models of
  social networks based on social distance attachment, Phys. Rev. E 70 (2004)
  056122.

\bibitem{mej4}
M.~E.~J. Newman, Fast algorithm for detecting community structure in networks,
  Phys. Rev. E 69 (2004) 066133.

\bibitem{aar}
A.~Clauset, M.~E.~J. Newman, C.~Moore, Finding community structure in very
  large networks, Phys. Rev. E 70 (2004) 066111.

\bibitem{and}
A.~Noack, R.~Rotta, Multi-level algorithms for modularity clustering, Lecture
  Notes in Computer Science 5526 (2009) 257--268.

\bibitem{gag}
G.~Agarwal, D.~Kempe, Modularity-maximizing graph communities via mathematical
  programming, Eur. Phys. J. B 66 (2008) 409--418.

\bibitem{jor}
J.~Duch, A.~Arenas, Community detection in complex networks using extremal
  optimization, Phys. Rev. E 72 (2005) 027104.

\bibitem{rog}
R.~Guimer\`a, L.~A.~N. Amaral, Functional cartography of complex metabolic
  networks, Nature 433 (2005) 895--900.

\bibitem{mej1}
M.~E.~J. Newman, Modularity and community structure in networks, Proc. Natl.
  Acad. Sci. USA 103 (2006) 8577--8582.

\bibitem{ame}
A.~Medus, G.~Acuna, C.~O. Dorso, Detection of community structures in networks
  via global optimization, Physica A 358 (2005) 593--604.

\bibitem{zhe}
Z.~Q. Ye, S.~N. Hu, J.~Yu, Adaptive clustering algorithm for community
  detection in complex networks, Phys. Rev. E 78 (2008) 046115.

\bibitem{gxu}
G.~Xu, S.~Tsoka, L.~G. Papageorgiou, Finding community structures in complex
  networks using mixed integer optimization, Eur. Phys. J. B 60 (2007)
  231--239.

\bibitem{are}
A.~Arenas, J.~Duch, A.~Fern\'andez, S.~G\'omez, Size reduction of complex
  networks preserving modularity, New J. Phys. 9 (2007) 176.

\bibitem{vin}
V.~D. Blondel, J.~L. Guillaume, R.~Lambiotte, E.~Lefebvre, Fast unfolding of
  communities in large networks, J. Stat. Mech. (2008) P10008.

\bibitem{ben}
B.~H. Good, Y.-A. de~Montjoye, A.~Clauset, The performance of modularity
  maximization in practical contexts (2009).
\newblock \href {http://arxiv.org/abs/0910.0165} {\path{arXiv:0910.0165}}.

\end{thebibliography}






% \end{thebibliography}

\end{document}